\documentclass[a4paper]{article}

\usepackage{INTERSPEECH2022}
\usepackage{color}
\usepackage{multirow}
\usepackage{url}
\usepackage[hyperfootnotes=false]{hyperref}
\usepackage{subfigure}
\usepackage{xcolor}
\usepackage{booktabs}

\newcommand\blfootnote[1]{%
  \begingroup
  \renewcommand\thefootnote{}\footnote{#1}%
  \addtocounter{footnote}{-1}%
  \endgroup
}

\title{Automatic Evaluation of Speaker Similarity}
\name{Kamil Deja$^{1*}$,Ariadna Sanchez$^2$, Julian Roth$^\dagger$, Marius Cotescu$^2$}
\address{
  $^1$Warsaw University of Technology, Poland\\
  $^2$Amazon Research, Cambridge, United Kingdom}
\email{kamil.deja@pw.edu.pl, \{cariadn, cotescu\}@amazon.com}

\begin{document}

\maketitle
\begin{abstract}
We introduce a new automatic evaluation method for speaker similarity assessment, that is consistent with human perceptual scores.
Modern neural text-to-speech models require a vast amount of clean training data, which is why many solutions switch from single speaker models to solutions trained on examples from many different speakers. Multi-speaker models bring new possibilities, such as a faster creation of new voices, but also a new problem - \emph{speaker leakage}, where the speaker identity of a synthesized example might not match those of the target speaker. Currently, the only way to discover this issue is through costly perceptual evaluations. In this work, we propose an automatic method for assessment of speaker similarity.
For that purpose, we extend the recent work on speaker verification systems and evaluate how different metrics and speaker embeddings models reflect Multiple Stimuli with Hidden Reference and Anchor (MUSHRA) scores. Our experiments show that we can train a model to predict speaker similarity MUSHRA scores from speaker embeddings with 0.96 accuracy and significant correlation up to 0.78 Pearson score at the utterance level.
\end{abstract}

\section{Introduction}

Recent development in neural text-to-speech (TTS) synthesis with models such as Tacotron~\cite{wang2017tacotron}, Transformer TTS~\cite{li2019neural} or FastSpeech~\cite{ren2019fastspeech} introduce high-quality synthesis for single-speaker TTS systems. However, in order to create a new voice, such architectures have to be retrained with a vast amount of clean recordings with the desired speaker characteristics. This is why there is a growing interest in TTS models supporting multiple speakers~\cite{gibiansky2017deep, hsu2018hierarchical,fan2015mutlispeaker,huybrechts2021low}, that additionally opens novel research opportunities such as speaker adaptation~\cite{taigman2017voiceloop}. \blfootnote{${}^*$ Work done during an internship at Amazon}\blfootnote{${}^\dagger$ Work done while at Amazon}

With new opportunities come new concerns. When TTS neural models are trained with recordings from more than one person, they might suffer from the problem of \emph{speaker leakage} -- a situation where the synthesized speech does not match the desired speaker identity. For example, when synthesizing speech with an input for a specific speaker, the model might discard this input and instead synthesize the output with speaker characteristics that are over-represented in similar training utterances. Currently, the only possibility to detect this issue is through perceptual evaluation and, since the problem might happen only for certain testcases, running them is time-consuming and costly. 

To facilitate development of new TTS systems, automatic objective evaluation methods emerged~\cite{kubichek1993mel,benesty2009pearson}. However, their performance is limited to verification of technical acoustic features of synthesized utterances. There are several machine learning based works, that try to automate different aspects of perceptual evaluation. E.g. in MOSNet~\cite{lo2019mosnet} authors train the model to predict Mean Opinion Scores on model naturalness reaching up to 0.67 Pearson correlation to perceptual scores on the  utterance level.  In this work, we investigate the problem of automatic speaker similarity assessment, and propose a solution that mimics the way in which naive listeners perceptually capture similarity between speakers. We base our solution on speaker features extracted with state-of-the-art speaker embedding models.

We first evaluate how the distance between speaker embeddings trained on a big corpus, without perceptual targets, correspond to the perception of speaker similarity. For this purpose, we evaluate several multidimensional metrics, and measure the correlation of distances between speaker embeddings and perceptual scores of speaker similarity according to the Multiple Stimuli with Hidden Reference and Anchor (MUSHRA) test~\cite{series2014method}.

Secondly, we investigate how we can improve this alignment with an additional regression model trained to predict a MUSHRA test score from extracted speaker embeddings. To achieve this, we propose a model based on multilayered perceptron. We experiment with several objectives to capture unequal distributions and uncertainty of scores from naive listeners, proposing a new loss function based on the mahalanobis distance~\cite{mahalanobis1936generalized}. Our experiments on a dataset with over 700 000 individual scores show a significant Pearson correlation of 0.78 at the utterance level between the proposed model's predictions and oracle perceptual scores.

\begin{figure}[t]
     \centering
     \includegraphics[width=0.99\linewidth]{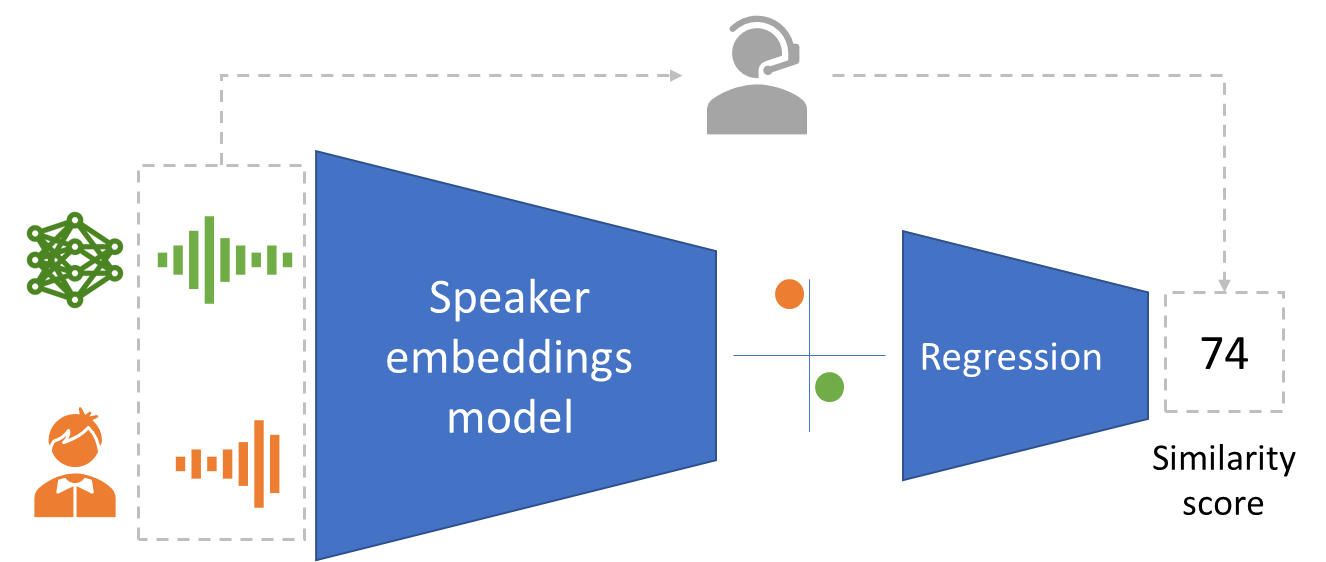}
     \caption{We propose a method for automatic evaluation of speaker similarity based on speaker embeddings model, trained to align with perceptual evaluation from naive listeners.}
     \label{fig:teaser}

\end{figure}

The main contributions of this work are: 
\begin{itemize}
\item Experimental verification of how speaker embeddings from currently used unsupervised methods correlate to the human perception of speaker similarity. 
\item Introduction of a novel method for automatic evaluation of speaker similarity.

\end{itemize}

\section{Related works}

\subsection{Speaker embeddings models}

Speaker similarity assessment highly resembles the problem of speaker verification, where the goal is to verify whether an audio sample is pronounced by a hypothesized speaker. 
In this work, we propose a solution built on top of a speaker embedding model trained in the same way as in a speaker verification problem. In~\cite{bai2021speaker}, authors survey recent deep learning approaches to this topic dividing them into two groups according to the training objective.

Models trained with classification objectives are usually based on the idea of binary classifier trained to predict whether a pair of inputs (either in time or time-frequency domain) comes from the same speaker. 
Different implementations of this technique uses different losses and activation functions, with variations of softmax and cross-entropy being the most common ones~\cite{richardson2015deep,huang2018angular}. Final speaker embedding is calculated as a latent representation on the classifier's hidden layer.

Since speaker verification or similarity assessment are an open set recognition task, the embedding space trained on a limited number of speakers is required to generalize well to unseen cases. In embedding-extraction based systems, to better align to this objective and minimize the within-class variance of embbedings, loss functions yield similarity scores from training trials. Similarity can be calculated in many ways, e.g. on pairs of examples with cosine~\cite{rahman2018attention} or contrastive loss~\cite{chung2018voxceleb2}, or three examples using triplet loss~\cite{li2017deep}. In GE2E~\cite{wan2018generalized} and~\cite{wang2019centroid}, authors introduce a generalized version of the embeddings extraction loss that, contrary to triplet loss, calculates the distance to the centroids of different speakers rather than random samples. In~\cite{saito2021perceptual} authors further develop this technique with additional input from perceptual speaker similarity matrix. 

\subsection{Automatic evaluation of TTS}

Several contemporary automatic evaluations of TTS models focus on technical acoustic features of synthesized outputs, that can be approximated with objective evaluations such as Mel-cepstral distortions (MCD)~\cite{kubichek1993mel} for comparison of voice spectrum and PCC~\cite{benesty2009pearson} or RMSE for prosody. Nevertheless in practical use-cases such as voice conversion problems (e.g. VCC Challenge~\cite{lorenzo2018voice}) objective evaluations are usually accompanied by perceptual ones~\cite{sisman2020overview}.

There are several works that try to automate perceptual assessment of TTS systems. The common idea is to use historical data to train the model in an end-to-end fashion to predict perceptual scores such as MOS. In several works~\cite{patton2016automos, lo2019mosnet}, authors experiment with different architectures trained to predict MOS naturalness evaluation scores of synthesized speech, reaching up to 0.67 Pearson correlation for utterance-level predictions. In~\cite{gamper2019intrusive, avila2019non}, authors evaluate similar approach for predicting the audio quality of synthetically distorted samples, reaching 0.87 Pearson correlation between predicted system scores and MOS. Similar solution for the problem of noise suppressors evaluation is proposed in~\cite{reddy2021dnsmos}. In SVSNet~\cite{hu2021svsnet} authors propose an extension to MOSNet to assessment of speaker similarity reaching up to 0.4 Pearson correlation on utterances from VCC2020 dataset, with further extension with x-vectors (0.79 Pearson correlation). 
In this work, we extend those approaches to the problem of speaker similarity assessment, using an additional speaker embedding model trained on large corpora without perceptual scores.  This approach limits our method to spectral features that might be extracted with such models.  

\section{Data}

We test proposed methods on historical evaluations of speaker similarity. Our dataset consists of 51 separate evaluation cycles for 13 target speakers and 354 systems, with total number of 18493 examples.  Each one was evaluated by at least 20 listeners, with a total number of 788 individual evaluators and 730~308 individual scores in total.  Listeners were composed of professional native speakers trained to perform this task. 
In those historical evaluation cycles, we compared different modern neural systems trained to mimic reference voices.  This includes,  among others, models created during development of methods for prosody transfer~\cite{karlapati2020copycat} or multi-speaker,  crosslingual synthesis~\cite{sanchez2022}.  Those acoustic models were trained with data of up to 130 different speakers in 8 languages.  The total training dataset size ranged between systems from 2 up to 580 hours. 
Each evaluation example consists of a pair of recordings -- source and reference that were synthesized with neural vocoders~\cite{jiao2021universal,  oord2018parallel}. 
Each listener scored each model between 0 and 100 for each utterance, where 100 means that the speakers are equal, and 0 that the speakers are completely different. Contrary to the original MUSHRA evaluation~\cite{series2014method}, we do not force users to score one system as 0 and 100.

\begin{figure}[t!]

		\centering
		\subfigure{\includegraphics[width=.49\linewidth]{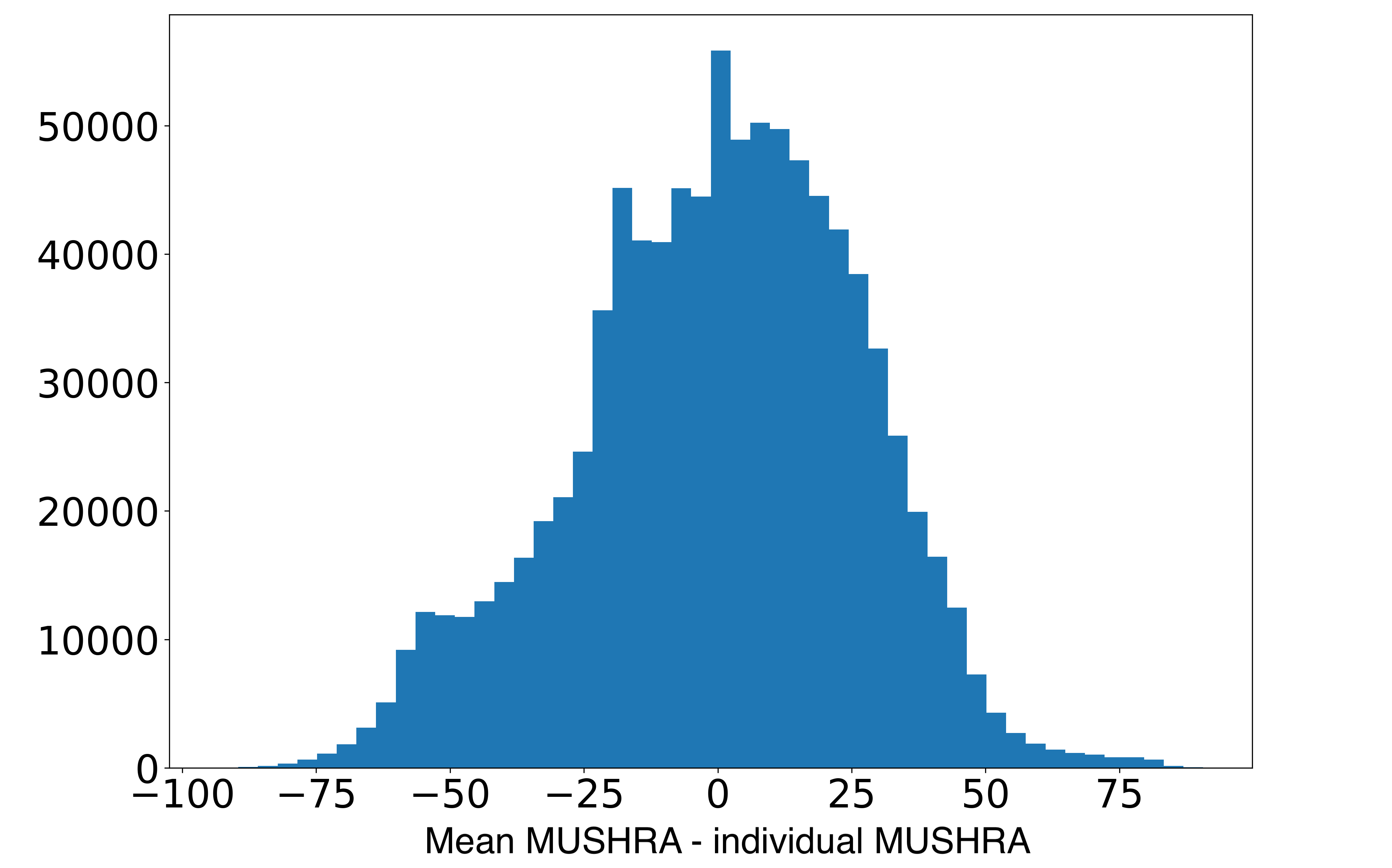}}
		\subfigure{\includegraphics[width=.49\linewidth]{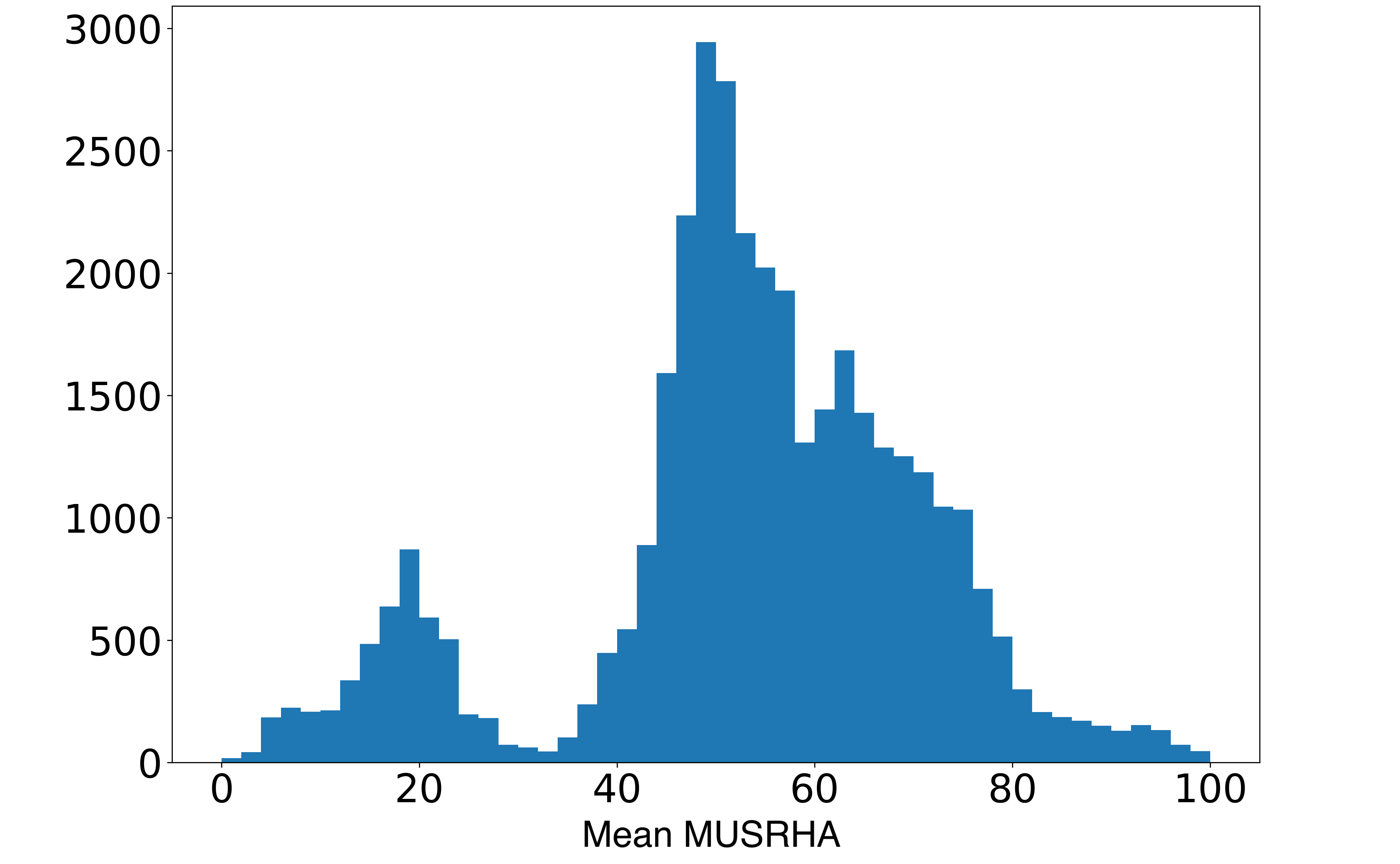}}

		\caption{\label{fig:data} Histogram of discrepancies between mean and individual scores for the same example (left) and distribution of MUSHRA scores in the dataset averaged over listeners scoring the same example (right).}
\end{figure}

We observe that the perception of speaker similarity varies considerably between listeners. In Fig.~\ref{fig:data}(left), we present a histogram of differences between mean and individual predictions. To reduce the variation from individual perceptual scores, in the final evaluation we compare predicted scores with MUSHRA scores averaged across all listeners for a given example. The distribution of such averaged scores is presented in Fig.~\ref{fig:data} (right). Discrepancies in human perception of speaker similarity as well as uneven distribution of scores with majority around 40-80 makes training of a predictive model challenging. Therefore in this work we propose loss functions that circumvent those problems. 

To approximate what is the highest quality we can expect from an automatic tool trained on top of such uncertain scores, we calculate the upper bound of proposed solution by checking how well listeners agree between each other. For that end, we randomly divide them into two groups and compare average MUSHRA scores for all samples between first and the second groups of listeners. With this comparison we reach Pearson correlation of $0.841 \pm 0.001$. In section ~\ref{sec:experiments} we examine proposed solutions with respect to this value.

\section{Method}
In this section, we introduce our method for automatic evaluation of speaker similarity. We build our solution on top of a pre-trained speaker embeddings model, which we use as an extractor of speaker characteristic related features.  In this work, we evaluate three types of backbone speaker embeddings model: a standard classification-based model proposed in~\cite{karlapati2020copycat}, a GE2E model~\cite{wan2018generalized} and a GE2E model with Glow normalization~\cite{valles2021improving}. Each one of them was trained with the same dataset, which is a combination of open Common Voice dataset~\cite{ardila2019common} (63\%) and in-house data with 581 hours of high quality studio recordings, and 562 hours of low quality recordings with background noise, out of which 393 were recorded by non-professional speakers. 
The full training set for speaker embeddings model consists of 39,880 speakers. During training, as a data augmentation step, we split examples into 1 second long sequences.

\subsection{Speaker embeddings models}
There are two main approaches in speaker embeddings training. The first one employs classifier trained to distinguish different speakers, while the second one directly calculates metrics between utterance representations calculated by the model. We compare which of those two strategies produce embeddings that better correspond to human perception by selecting an exemplar backbone model from each group.

As a representative of the first group we use a model proposed in~\cite{karlapati2020copycat}. It takes Mel-spectrogram as an input and provides speaker embeddings on the utterance-level, as hidden layer representations of the speaker classifier. 
Part of the network responsible for generating embeddings uses 2D convolutional layers to reduce the dimensions along both the time and frequency axes followed by a GRU layer along the shortened time axis. Auxiliary similarity network, on top of the embeddings extractor, consists of two fully connected linear layers and ReLU activation. The model is trained with Binary Cross Entropy loss to predict whether two examples came from the same speaker. 

For the second type of speaker embeddings model we propose a similar architecture trained with GE2E loss function~\cite{wan2018generalized}.  The main idea of this approach is to train a model that creates embeddings directly minimizing the distance between examples from the same speaker while, at the same time, maximizing the distance to the embeddings from other speakers.  In GE2E, authors propose to calculate distances between an anchor and centroids of a given speaker embeddings rather than individual random samples. Because of the direct metric training, in this approach we do not have an auxiliary classifier.

Finally, we follow the idea introduced in~\cite{valles2021improving} and explore whether additional normalization of speaker embeddings through a Glow model~\cite{kingma2018glow} improves the correlation of distances in the new normalized latent space to the perceptual speaker similarity scores. To that end, we process embeddings from the best performing GE2E model through a Glow model trained to attain Gaussian distribution on the output latent space. We use the same model architecture as in~\cite{valles2021improving}, which samples from the Gaussian distribution defined by the normalized embeddings, both at training and inference time.  In baseline solution we directly calculate distances between speaker embeddings produced by a given backbone model and compare them with perceptual scores of speaker similarity.

\subsection{Regression}
In order to improve the alignment between speaker embeddings and perceptual MUSHRA scores, we build an additional regression model that predicts MUSHRA speaker similarity scores from speaker embeddings of two utterances. We experiment with a simple model that has two fully connected layers, LeakyReLU activation and dropout. We train the model with whole utterances from the evaluation dataset in cross-validation schema.  As a baseline, we use Mean Squared Error (MSE) loss. However, as presented in Fig.~\ref{fig:data} (right), the distribution of similarity scores is not uniform, with the majority of examples scoring between 40 and 80. To mitigate this discrepancy, we weight the MSE loss with weights that are inversely proportional to the density of the target scores.

Apart from this unequal distribution, we can also observe significant discrepancies in individual scores between different listeners for the same example, as presented in Fig.~\ref{fig:data} (left). Therefore, we also experiment with other loss functions that incorporate information about the distribution of the target variable, instead of a single average point. For this purpose, we propose a loss function based on the Mahalanobis distance. For each training example, we calculate the mean and standard deviation of scores distribution. Then, as a training loss, we measure how many standard deviations there are between the predicted value and the oracle mean MUSHRA score. The exact loss function for one data example is noted in Eq.~\ref{eq:loss}.

\begin{equation}
\mathcal{L} = \frac{|E(X) - \hat{X}|}{\sigma(X)}
\label{eq:loss}
\end{equation}

\noindent where X is the random variable of scores for a given example and $\hat{X}$ is a model's prediction.

The proposed mahalanobis distance-based loss amplifies differences in examples that have a better alignment in perceptual evaluation -- where listeners are more certain about the similarity of two speakers -- while reducing the penalty for wrong scores on perceptually unclear ones. 

To even better capture the distribution of target scores, we also experiment with a model trained on raw individual scores from all listeners, without previous aggregation. This leads to a situation where, in each epoch, the model is trained several times with exactly the same input (embeddings of source and target recordings), but with a variety of target scores from different listeners. We present results of those experiments in the next section.

\begin{table}[t!]
  \caption{Pearson correlation of distances between speaker embeddings and averaged speaker similarity MUSHRA scores}
  \label{tab:metric_results}
  \centering
  \begin{tabular}{ r | l  r }
    \toprule
    \textbf{Embeddings model} & \textbf{Metric} & \textbf{Pearson correlation}\\
    \midrule
    \multirow{2}{*}{classification-based}& Euclidean 	& -0.52\\
                           	  	   		& Cosine 		& -0.51\\
    \multirow{2}{*}{\textbf{GE2E}} 	& Euclidean  	& \textbf{-0.61} \\
                             			  	& Cosine 		& \textbf{-0.63}\\
    \multirow{2}{*}{GE2E +Glow}		& Euclidean 	& -0.42\\
                              				& Cosine 		& -0.31\\
    \bottomrule
  \end{tabular}
  
\end{table}
\begin{table}[t!]
  \caption{Comparison between different loss functions of the regression model on GE2E speaker embedding. }
  \label{tab:regression_loss_results}
  \small
  \setlength{\tabcolsep}{3pt}
  \centering
  \begin{tabular}{ r | c c c c}
    \toprule
    \multirow{2}{*}{\textbf{Loss function}} & \textbf{Pearson} &\textbf{Avg. Pearson}& \multirow{2}{*}{\textbf{Accuracy}} & \multirow{2}{*}{\textbf{RMSE}}\\
    &\textbf{correlation} &\textbf{correlation}&&\\
    \midrule
    \multirow{1}{*}{MSE}      & 0.69 & $0.67\pm 0.1$  & 0.93 & 12.36\\
    \multirow{1}{*}{Weighted MSE}     	  & 0.71 & $0.69\pm 0.12$ & 0.93 & 12.33\\
    \multirow{1}{*}{Mahalanobis}   & 0.71 & $0.69\pm 0.11$  & 0.94 & \textbf{ 12.12}\\
    \multirow{1}{*}{\textbf{Mahalanobis}}   &\multirow{2}{*}{\textbf{ 0.78 }}& \multirow{2}{*}{$\mathbf{0.71\pm 0.145}$ }& \multirow{2}{*}{\textbf{0.96}}& \multirow{2}{*}{\textbf{12.04}}\\
    \textbf{single}&&&&\\
    \midrule
    \multirow{1}{*}{Upper bound} &0.84&-&0.97&7.68\\ 
    \bottomrule
  \end{tabular}
  
\end{table}

\begin{table}[t!]
  \caption{Comparison between original and predicted MUSHRA scores for regression model trained with Mahalanobis loss function and different base speaker embeddings model.}
  \label{tab:regression_results}
  \small
  \setlength{\tabcolsep}{3pt}
  \centering
  \begin{tabular}{ r | c c c c}
    \toprule
    \multirow{1}{*}{\textbf{Embeddings}} & \textbf{Pearson} &\textbf{Avg. Pearson}& \multirow{2}{*}{\textbf{Accuracy}} & \multirow{2}{*}{\textbf{RMSE}}\\
    \multirow{1}{*}{ \textbf{model}}&\textbf{correlation} &\textbf{correlation}&&\\
    \midrule
    \multirow{1}{*}{Classification}      & 0.77 & $\mathbf{0.71\pm 0.14}$ &\textbf{ 0.96} & \textbf{11.96}\\
    \multirow{1}{*}{\textbf{GE2E}}     	  & \textbf{0.78} & $\mathbf{0.71\pm 0.145}$ & \textbf{0.96 }& \textbf{12.04}\\
    \multirow{1}{*}{GE2E +Glow}   & 0.72 & $0.67\pm 0.18$  & 0.94 & 13.11\\
    \midrule
    \multirow{1}{*}{Upper bound} &0.84&-&0.97&7.68\\ 
    \bottomrule
  \end{tabular}
  
\end{table}

\section{Experiments \label{sec:experiments}}

In this section, we show the results of the experiments that evaluate the proposed solutions. Our main evaluation criteria is the Pearson correlation between predictions and averaged MUSHRA scores at the utterance level. 

\subsection{Evaluation results}
We first compare whether multidimensional metrics between speaker embeddings of source and reference recordings correlate with the speaker similarity perceptual score. In Tab.~\ref{tab:metric_results}, we show results of utterance level correlations between Euclidean or cosine distance and averaged MUSRHA evaluation scores on the whole dataset. We also analyze results from the three backbone speaker embedding models, which shows how GE2E based embeddings outperform other models. 
This is caused by the fact that GE2E loss function explicitly forces a model to group speakers according to their cosine distance.  Having an additional Glow normalization on top of it undermines those meaningful representations.

\begin{figure}[t!]
     \centering
     \includegraphics[width=0.95\linewidth]{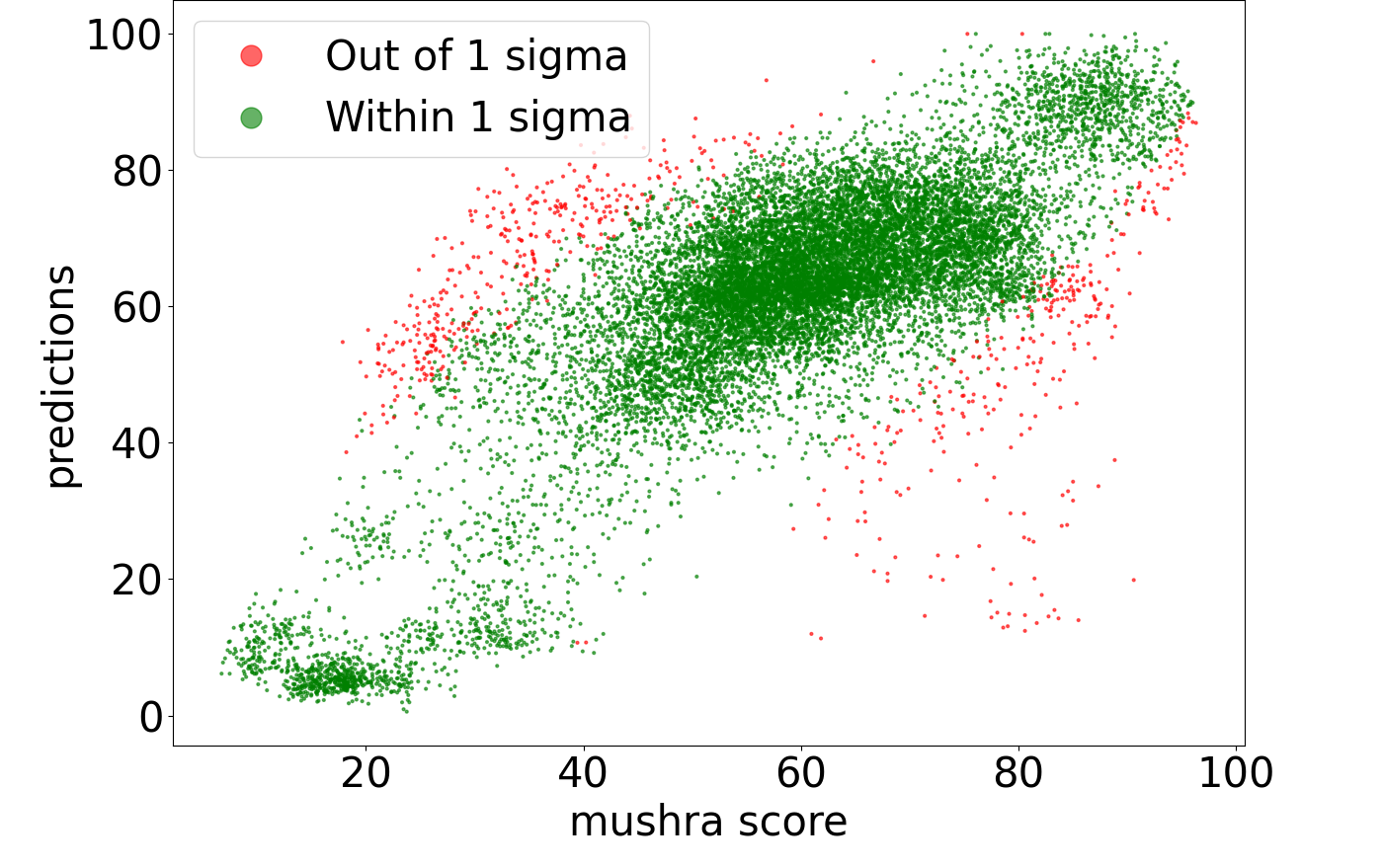}
     \caption{Scatter plot comparison between predictions and mean MUSHRA scores. Predicted MUSHRA scores are well aligned to original averaged scores, with discrepancies that are mostly consistent with those observed in the original data, and almost 96\% of examples within $1\sigma$ from the mean score.}
     \label{fig:accuracy}

\end{figure}

To improve the alignment between speaker embeddings, we propose a regression model that is trained to predict perceptual MUSHRA scores. We train the model using a cross-validation schema with 10 splits, and present the correlation on joint testsets (Pearson correlation) and averaged correlation scores calculated separately on each testset (Avg. Pearson correlation). Additionally, we measure the testset's Root Mean Squared Error (RMSE) of predictions and accuracy. For the latter, as true positives we consider predictions within one standard  deviation from the mean perceptual score. 
In Tab.~\ref{tab:regression_loss_results}, we compare scores for the regression model trained with different loss functions and GE2E as the backbone speaker embedding model. Experiments indicate that Mahalanobis based loss function indeed improves performance of the model yielding an even lower RMSE than the model trained explicitly with MSE as a loss function. Training on individual scores (Mahalanobis single) allows the model to further improve how it captures the distribution of scores, which results in a correlation score of 0.78 slowly approaching the upper bound of 0.84. In figure Fig.~\ref{fig:accuracy}, we present a scatter plot comparing the predictions with the mean perceptual MUSHRA scores for the best performing Mahalanobis loss and trained on individual scores. In the picture, we highlight wrong predictions -- examples where the predicted values are off by at least one standard deviation from the mean calculated from distribution of individual scores for a given example.

In Tab.~\ref{tab:regression_results}, we overview the results of the regression model trained with the best performing loss function, but using a different backbone speaker embedding model. With an additional simple neural model, we can improve the alignment to perceptual MUSHRA scores for all speaker embeddings including classification-based ones, which is no longer significantly worse then GE2E. 
Nevertheless, additional Glow normalization of speaker embeddings still results in lower correlation.

\subsection{Sub-utterance performance}

Finally, since speaker leakage may happen also in the part of the utterance, we evaluate the performance of the proposed method at the sub-utterance level. 
 For this purpose, we prepare a dataset by splitting all spectrograms into one second long pieces. 
We assume that for data samples in our evaluation cycles MUSHRA scores were constant along the whole utterance, therefore for
each pair of new examples 
, we assign the same MUSHRA score as for the whole utterance, assuming that each score assigned by listeners is constant. We train the best performing regression model with Mahalanobis loss function on one second long pieces and individual scores. In this experiment, we observe almost as good performance as for the utterance level model, achieving a Pearson correlation of $0.76 \pm 0.06$ and accuracy equal to 0.95.

\section{Conclusions}
To our knowledge, in this work we introduce the first automatic evaluation method for speaker similarity assessment. 
We experimentally show that latent spaces of speaker embedding models trained without access to perceptually labeled data  already share some similarities to what is perceived by humans, and  propose a regression model to better align predictions to perceptual evaluations of speaker similarity. 
In our experiments with real evaluation cycles composed by 730 308 individual scores we achieve up to 0.78 Pearson correlation of predicted values to averaged perceptual MUSHRA scores, and up to 0.96 accuracy. Those results approach the upper bound of a solution (0.84) indicating that proposed method may be used in practice to facilitate process of multispeaker models evaluation and hence  accelerating the development of novel TTS methods. Furthermore, the direct combination of speaker similarity automatic evaluation method with speaker embeddings or multispeaker models as proposed in~\cite{manocha2020differentiable} might lead to the development of novel, more reliable methods. 

\bibliographystyle{IEEEtran}

\bibliography{mybib.bib}

\end{document}